\documentclass[aps,prl,twocolumn,superscriptaddress]{revtex4-1}
\usepackage{graphicx}
\usepackage{setspace}
\usepackage{amsmath}
\usepackage{amssymb}
\usepackage{color}
\usepackage{array}
\usepackage{subfigure}
\usepackage{hyperref}
\usepackage{float}
\begin{document}
\title{Tunable lattice Induced Opacity in atom-lattice interaction}
\author{Chen Zhang}
\affiliation{Department of Physics and JILA, University of Colorado, Boulder, Colorado 80309-0440, USA}
\affiliation{Department of Physics, Purdue University, West Lafayette, Indiana 47906, USA}\author{Chris H. Greene}
\affiliation{Department of Physics, Purdue University, West Lafayette, Indiana 47906, USA}
\date{\today}
\begin{abstract}
A lattice-induced opacity is identified in the scattering process of a normally-incident matter wave from a two dimensional lattice of atoms. This system can be treated as an analogue of a confinement induced resonance.  Specifically by modifying the s-wave scattering length between atoms in the incident matter wave and the lattice-confined atoms, the transmission of the matter wave can be tuned controllably from zero to one. Based on this, we propose a possible scheme for a matter wave transistor. When the transmission of the matter wave is tuned to zero, a matter wave cavity can be formed by placing  two such lattice planes of atoms parallel to one another. At higher kinetic energies, the two dimensional lattice of atoms can also serve as a matter wave beam splitter and a wave plate.  
\end{abstract}

\maketitle
%\section{Introduction}
Cold matter waves used as current carriers and as probe beams have become a new area of quantum control discussed in many theoretical proposals \cite{PhysRevLett.103.140405,PhysRevLett.101.265302,PhysRevLett.93.140408,PhysRevA.84.013608}. In contrast to electrons, ultra cold matter waves are more coherent, and they propagate in a much cleaner environment, which allows study of an idealized system from which all inessential complications are stripped. In particular, ultracold matter waves exhibit novel coherence and transport properties \cite{PhysRevLett.64.1658,PhysRevLett.82.3008,PhysRevLett.78.582}. Atomtronics, proposed and realized in an ultracold atomic gas \cite{nature379382, arxiv1208.3109}, is analogous to electronics in a solid-state system. Quantum diodes and transistors are the basic knobs that control the propagation of an ultracold matter wave in optically or magnetically trapped atomic ensembles \cite{arxiv1208.3109, Chen16082013}. They behave like the PN junctions and classical transistors that control currents in semiconductors. The additional tunability of the interactions between ultracold atoms that originates from the rich internal degrees of freedom in atoms, which contrasts with the electron-photon interaction in a solid-state system. This tunability is realized through external electric, magnetic, or optical fields e.g., ramped across a Fano-Feshbach resonance\cite{RevModPhys.82.1225}. Matter wave-scattering experiments were proposed to probe the properties of atomic ensembles,  by extracting atom-atom correlation functions from the differential-scattering cross sections in experiments that scatters an atomic beam from an ultracold cloud of atoms \cite{PhysRevA.60.R769, PhysRevLett.105.035301}. 
\par
The manipulation of ultra cold matter wave started right after the realization of laser cooling of atoms \cite{PhysRevLett.82.3008, PhysRevLett.64.1658}. Current experiments are actively exploring the use of ultracold atomic matter waves as beam sources to probe novel phases in optical lattices \cite{PhysRevA.87.013623, Gadwaynp2012} . For example, Gadway \textit{et. al} \cite{Gadwaynp2012} determined the spatial ordering of a one-dimensional Mott insulator. This experimental advance opens the door to the use of these techniques to probe novel phases of ultracold atomic ensembles, in analogy to optical Bragg spectroscopy and ARPES (angle resolved photon emission spectroscopy) \cite{PhysRevLett.82.4569,PhysRevLett.91.150404,PhysRevLett.106.215301}.  This experimental technique is complementary to the time of flight and in-situ measurements typically used to probe the properties of ultracold atomic gases in current generation experiments. However these early explorations have treated the atom-atom interactions in the far-off-resonance regime. 
\par
The tunability of the atom-atom interaction is essential for the coherent control of an atomic matter wave, as has already been demonstrated in a unitary Bose gas \cite{arxiv1308.3696} and in a strongly interacting Fermi gas \cite{Regal2003}. For an ultracold system in reduced dimensions, as was first discovered by Olshanii \cite{Olshanii1998} and later generalized by Blume \textit{et. al} \cite{Blume2004} and Kim \textit{et. al} \cite{PhysRevA.72.042711}, atomic scattering processes can exhibit resonance features without entering the 3D strongly interacting regime, thanks to the interplay between interparticle interaction and confinement geometry. Such resonances provide a key opportunity to make an unusual type of matter wave transistor. As we will show below, a ``slow\rq{}\rq{} beam is more favorable for the purpose of quantum control of an atomtronic transistor in reduced dimensions, in contrast with a traditional Bragg spectroscopy experiment where ``fast\rq{}\rq{} electrons or neutrons are needed to avoid multiple scattering events that complicate the dispersion relation mapping. 
\par
\begin{figure}
\includegraphics[angle=0,scale=0.3]{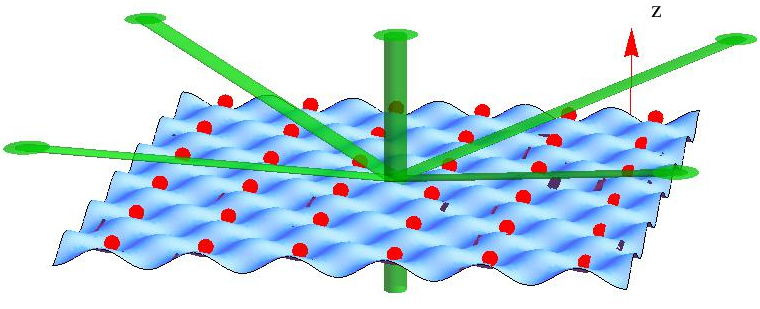}
\caption{(color online) Sketch of a beam of an atomic matter wave scattering from an infinite 2D square lattice of fixed atoms. For simplicity, the incoming wave is restricted to be normal to the 2D optical lattice.}
\label{TwoDLatticeAndWave}
\end{figure}
\begin{figure}
\includegraphics[scale=0.3]{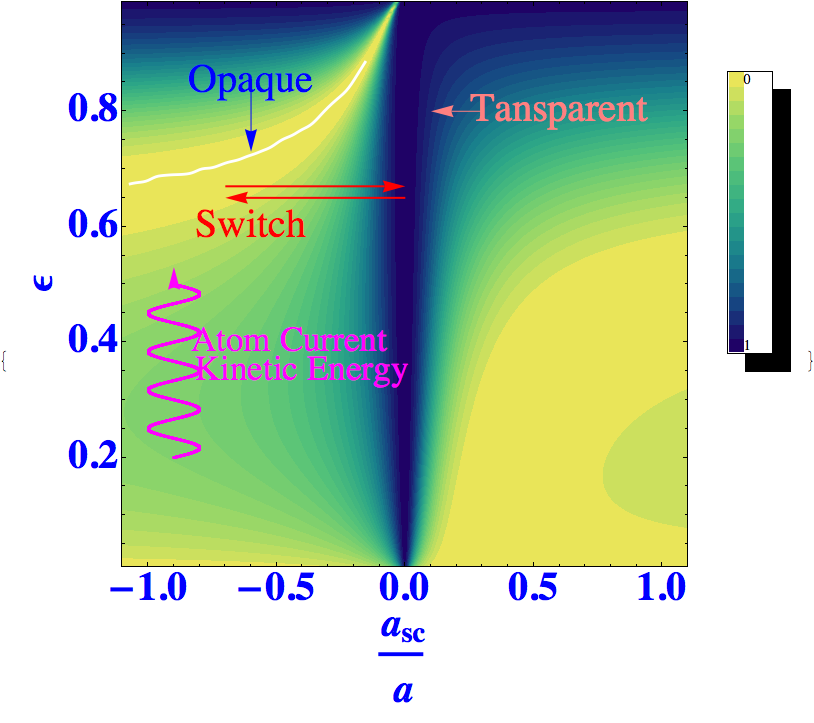}
\caption{(color online) This graph plots the quasi-1D transmission probability, $|1+f_{0,0}|^2$, as a function of the 3D scattering length $a_{sc}$ and the kinetic energy $\epsilon$ of the incoming atom for scattering in the case of a single open channel. There are two regions in this parameter space where the scattering amplitude is close to zero. However, only the one that is associated with a negative 3D scattering length $a_{sc}$ provides a usable lattice-size-induced opacity. At positive values of $a_{sc}$, although there exists a region where $a_{sc}$ and $\epsilon$ cooperatively result in a small transmission amplitude, the tunability there is poorer than the negative $a_{sc}$ region. This is because in the region $a_{sc}>0$, even tuning $\epsilon$ and $a_{sc}$ over a large range does not significantly change the transmission amplitude.}
\label{t00}
\end{figure}

This \textit{Letter} begins with a discussion of an atomic beam that scatters from an infinite 2D atomic lattice, as shown in Fig. \ref{TwoDLatticeAndWave}. A quantum matter wave transistor mechanism is proposed that is based on the transmission-reflection property of a coherent matter wave. This mechanism relies on controlling the atom-atom interactions in a reduced dimension scattering process that is shown in Fig. \ref{t00}. In addition, a possible mechanism for coherently controlling the peak scattering intensity into the angles that correspond to different Bravais lattice vectors is discussed, in the framework of a series of Fano-Feshbach resonances. The proposed mechanism could be a promising candidate for an atomtronic transistor using ultracold atoms confined in a 2D optical lattice at an ultracold temperature.
\par
%\section{Lattice size induced opacity and atomtronic transistor}
The Hamiltonian that describes the atomic matter wave interacting with a 2D lattice of tight-trapped scattering centers reduces to:
\begin{equation}
H=-\dfrac{\hbar^2\nabla^2}{2\mu}+\sum_{(i_x, i_y)}V(\mathbf{r}-\mathbf{R}_{i_x, i_y}),
\end{equation}
in which $\mathbf{R}_{i_x, i_y}=a(i_x,j_y,0), i_x,j_y \in \it{Z}$ represent the positions of the scatterers, and $a$ denotes the square lattice constant.
Each scattering potential is approximated by a regularized delta function $V(\mathbf{r})=\frac{2\pi a_{sc}}{\mu}\delta(\mathbf{r})\frac{\partial}{\partial r}(r\cdot)$, where $a_{sc}$ is the $s$-wave scattering length between a beam atom and each scatterer.
\par
For a ``slow\rq\rq{} atomic beam, in which only the lowest transverse channel of the 2D lattice is energetically open, the quasi-1D scattering amplitude $f_{0,0}$ in the lowest channel $(0,0)$ is calculated analytically and regularized using the zeta-function regularization method developed in our preceding work \cite{PhysRevA.88.012715}. (We denote the reflection amplitude as $f_{0,0}$, while the transmission amplitude is $1+f_{0,0}$.) The asymptotic wave function for the scattered atomic matter wave is:
\begin{equation}
\begin{array}{lll}
\Psi=e^{ikz}+f_{0,0} e^{ik|z|}\\
\\
+\displaystyle{\sum_{(n_x,n_y)\ne (0,0)}} \exp(-\dfrac{2\pi}{a}\sqrt{n_x^2+n_y^2-\epsilon})e^{in_x x+in_y y} C_{n_x, n_y},
\end{array}
\end{equation}
where $k$ is the momentum of the incoming matter wave in the $z$ direction, the corresponding kinetic energy in the lattice recoil energy unit is $\epsilon=\frac{1}{E_{0}}\frac{\hbar^2 k^2}{2\mu}$, $E_0=\frac{\hbar^2}{2\mu}(\frac{2\pi}{a})^2$. A Green\rq{}s function calculation for a 3D optical lattice was carried out in \cite{PhysRevLett.92.080401}. Because we consider only a normal incoming matter wave to the lattice plane, the Green\rq{}s function formulation is simplified to a coupled-channel formulation. The scattering amplitude encapsulates contributions from all the closed channels, turning out to be an Epstein zeta function \cite{e1994zeta}:
\begin{equation}
f_{0,0}=\dfrac{2\pi a_{sc}}{a^2 ik}\dfrac{1}{1-\dfrac{2\pi a_{sc}}{a^2 ik}+\dfrac{a_{sc}}{a}\Lambda_{E}(\epsilon)},
\end{equation}
where 
\begin{equation}
\small{
\Lambda_{E}(\epsilon)=\dfrac{\partial}{\partial z} \left.\left(z\sum_{(n_x,n_y)\ne (0,0)}\dfrac{\exp(-\dfrac{2\pi|z|}{a}\sqrt{n_x^2+n_y^2-\epsilon})}{\sqrt{n_x^2+n_y^2-\epsilon}}\right)\right|_{z\to 0}.
}
\end{equation}
The positions of resonances correspond to zeros of the transmission amplitude $(1+f_{0,0})$, which gives $\frac{a_{sc}}{a}=-\frac{1}{\Lambda_{E}(\epsilon)}$ at resonance. At such a resonance there is vanishing transmission probability, or in other words, ``opacity\rq{}\rq{} of the 2D lattice to the matter wave.  
\par
The mapping from the reduced dimensional system to a transistor works as follows. In a transistor, the crux of the idea is to use a small tunable voltage to control a comparatively large current. The atomic matter wave, of course, serves as the current from the emitter that is experimentally implemented using two atomic reservoirs with different chemical potentials. The tunable scattering length $a_{sc}$ plays the role of the ``control voltage\rq{}\rq{}. It is not literally a voltage, but it exhibits every aspect that a ``control voltage\rq{}\rq{} element in an atomtronic circuit needs to have. This inter-species scattering length can be tuned either very adiabatically or else quite diabatically by controlling the ramping speed of the external magnetic field, which makes this reduced dimensional system a candidate for gate operations. The energies where a resonance occurs lie between the lowest and the first-excited transverse mode(s) of the 2D lattice plane. This resonance happens at a nondiverging value of the 3D inter-species scattering length $a_{sc}$, keeping the system from having strong three-body losses. 
\par
The fact that the numerator of $1+f_{0,0}$ is purely real makes it possible to find a complete ``shut down\rq{}\rq{} of the transmitted matter wave. This corresponds to one case of the Fano-Feshbach theory, namely, the case in which a single bound state is embedded in a single continuum. However if there are more than one continuum channels present, the transmitted wave intensity can no longer be fully turned off, but only tuned to a local minimum. Consequently, the signal-to-noise ratio of the transparency and opacity of the lowest open channel deteriorates in the presence of multiple open channels. 
\par
Besides proposing the matter wave transistor scheme, we also estimate the temperature range in which this matter wave transistor can work perfectly, meaning that the velocity spread in the both the transverse direction and the longitudinal direction is comparatively small to the recoil velocity of the optical lattice $v=\frac{2\pi\hbar}{mL}$. Continuous atomic current from a BEC was demonstrated in output coupler \cite{PhysRevLett.78.582}, in which the velocity of the atomic current is adjusted by the rf frequency. A typical optical lattice constant ranges from $1.3\mu$m to $9.1\mu$m \cite{PhysRevLett.110.153001}, and an even smaller value has recently been achieved using a magnetic field and a type II superconductor \cite{PhysRevLett.111.145304}. And the finite temperature will result in velocity spread in the atomic beam $\Delta v\sim\sqrt{2k_B T/m}$ \cite{PhysRevLett.64.1658, PhysRevLett.78.582}, $m$ is the atomic mass. The spread of the velocity and the characteristic lattice velocity scale different with atomic mass. To achieve small relative spread, lighter atomic spices are favored. For example, in the recent sub wavelength optical lattice \cite{PhysRevLett.111.145304}, $\Delta v/v$ can be as small as 3\% for lithium BEC. However, the atomic velocity spread $\Delta v=\sqrt{2k_B T/m}$ is an estimation of the upper bound. The noise in the longitudinal velocity spread were reduced to 1/10 of this upper bound in by Bloch \textit{et.al} in \cite{PhysRevLett.78.582}. So the lattice indued opacity can also be achieve in traditional optical lattice with smaller recoil energy than sub wave length optical lattice.
\par
%\section{Controlling the intensity of Bragg peaks}
The discussion above implies that, for the quantum transistor proposed here, the temperature criterion is determined by the kinetic energy of the incoming matter wave and its spread ,such that only the lowest channel in the 2D periodic lattice plane is energetically open. In multi-open-channel quasi-1D scattering, no zero point of transmission in any single channel occurs, meaning no complete opacity of any channel can be achieved. However, in the multi-open-channel scenario, a similar confinement-induced opacity can be observed while the imaginary part of the complex transmission amplitude hits zero. The scattering amplitude in a system having multiple open channels was explained and classified by Fano \cite{PhysRev.124.1866} as an example of multi-continuum resonances, nowadays termed a Fano 
lineshape. 
\par
After discussing the single-open-channel case, a higher energy incoming atomic beam is considered. The discussion below assumes that the incident channel is the lowest transverse mode of the 2D lattice plane $\psi(z\to-\infty)=\psi_{0,0}(x,y)e^{ikz}$. The reason this case is singled out is that this matter wave in the lowest mode is the easiest to implement experimentally. By increasing the $k_z$ in the incoming wave, it is possible to excite higher modes in the $xy$-direction as outgoing waves. We now demonstrate the inelastic scattering in a problem involving two open channels of the 2D square lattice. Under the assumption that no modes in the excited channels exist in the incoming beam, the general expression of the outgoing matter wave to a higher transverse mode is:
%\begin{equation}
%\small{1+f_{0,0}=\dfrac{1+\dfrac{a_{sc}}{a}\Lambda_{E}-\dfrac{a_{sc}}{a}\sum_{(n_x, n_y)\ne(0,0)}g(E_n)\dfrac{2\pi}{ai q_{n}}}{1+\dfrac{a_{sc}}{a}\Lambda_{E}-\dfrac{a_{sc}}{a}\dfrac{2\pi}{aik}-\dfrac{a_{sc}}{a}\sum_{(n_x,n_y)\ne(0,0)}g(E_n)\dfrac{2\pi}{ai q_{n}}}},
%\end{equation}
\begin{equation}
f_{0,\pm 1}=f_{\pm 1, 0}=s_{0,0}\dfrac{\frac{2\pi}{a i q}\frac{a_{sc}}{a}}{1+\frac{a_{sc}}{a}\Lambda_{E}-\frac{2\pi}{aik}\frac{a_{sc}}{a}-4\frac{2\pi}{aiq}\frac{a_{sc}}{a}}.
\end{equation}
In multi-open-channel scattering, as an analogue to the confinement-induced resonance in the single-open-channel case, the zero point of $1+\frac{a_{sc}}{a}\Lambda_{E}$ causes a $\pi$ phase jump of the outgoing wave into $(0,\pm 1)$ and $(\pm 1, 0)$ with respect to the incoming wave. Moreover, the zero point of $1+\frac{a_{sc}}{a}\Lambda_{E}$ gives rise to a maximum inelastic scattering amplitude. Thus for a given incoming wave vector perpendicular to the 2D lattice plane, a similar resonance can also emerge in the inelastic scattering process by tuning the 3D scattering length. This process can be viewed as a coherent matter wave beam splitter, in the sense that it creates phase-coherent waves that propagate in different spatial directions. Therefore, this process is a class of the Kapitza-Dirac type thin-grating diffraction, which was first realized in BEC scattering from a 2D optical lattice \cite{Guptapaper}, in which BEC atoms directly gain momentum from the counter-propagating laser that forms the optical lattice. However, the configuration in this letter is different from that in the approach of Gupta \textit{et.al} \cite{Guptapaper}. In the present reduced-dimensional system, the number of interference peaks is controlled by the incoming matter wave kinetic energy and the short range interaction between atoms in the matter wave and atoms trapped in the 2D optical lattice. As a result, it is possible to completely turn off the intensity of the perpendicular wave which cannot occur in a traditional Kapiza-Dirac thin grating.
\par
\begin{figure}
\centering
\includegraphics[scale=0.5]{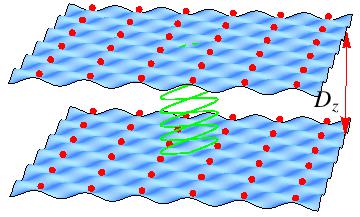}
\caption{(color online) Sketch of a matter wave cavity utilizing two parallel 2D optical lattices where the effective 1D interaction strength $g_{1D}$ for each of them is infinite.}
\label{CavityandWave}
\end{figure}

Since the $s$-wave interaction between atoms in the matter wave and those in the 2D lattice of atoms can be described using a single effective 1D delta function in $z$: $g_{1D}\delta(z)$, it is possible to make a matter wave cavity by placing two optical lattices separated by a certain distance $D_z$, sketched in Fig. \ref{CavityandWave}. The quasi-1D matter wave cavity can be simply described by two  delta function potentials in 1D:
\begin{equation}
H_{1D}=-\dfrac{\hbar^2}{2\mu}\dfrac{\partial^2}{\partial z^2}+g_{1D}\delta(z-\frac{D_z}{2})+g_{1D}\delta(z+\frac{D_z}{2}).
\end{equation}
The cavity is realized by tuning the transmission coefficients of the 2D optical lattices outside the region $(-\frac{D_z}{2},\frac{D_z}{2})$ to zero. By adjusting the distance between the two lattice planes, the standing matter wave ($k_zD_z=n\pi, n=1,2,\cdots$) can be created inside this cavity configuration.
This matter wave resonant cavity scheme can potentially be applied in matter wave interferometry experiments. The high precision and controllability of atoms in matter wave interferometry should enable the combination of matter waves and optical lattices to be implemented as an important precision measurements platform \cite{PhysRevLett.106.038501,PhysRevA.85.013639}. 
\par
%\begin{figure}
%\includegraphics[scale=0.3]{ManyOpenChannel00.png}
%\caption{(color online) The elastic scattering amplitude into the channel $(0,0)$ as a function $a_{sc}$ and incoming beam\rq{}s total energy when the amplitudes in the excited channels are zero. The brighter color corresponds to a small value of elastic scattering amplitude into $(0,0)$ channel. In the presence of more open channels, the elastic scattering amplitude is diluted and will never vanish, and the corresponding Fano-Feshbach resonance is getting weaker. This figure is plotted in the same color scale as Fig. \ref{t00}}.
%\label{ManyOpenChannel00}
%\end{figure}
%\section{Looking for Fano-Feshbach resonance in quasi-1D system}
The quasi-1D scattering process with a transverse confinement does, in principle, apply to Fano-Feshbach resonance theory, but the resonance width is quite significant compared to the assumption of an isolated resonance that was made in \cite{PhysRev.124.1866}. This large width means each individual resonance profile may deviate from the exact analytical expression derived by Fano because of an influence from a series of cuts by transverse thresholds. However, the analytical expression still preserves the asymmetric feature compared to a symmetric Lorentz line shape. Quantitatively, this asymmetric feature is embedded in the piecewise Epstein zeta function for different energy regions $E\in(E_{0}, E_{1}), E\in(E_{1}, E_{2})\cdots$. 
\par
In the expression of the Epstein zeta function, when the energy reaches each channel threshold $E_{n}$, the divergent scattering cross section behaves like $\sim\frac{1}{\sqrt{E-E_{n}}}$, while other terms vary much more slowly. This diverging behavior reflects the fundamental origin of Fano-Feshbach resonance, namely, the embedded bound state in a continuum. But the actual lineshape of observables is changed because the near-threshold density of states is not flat as was assumed in the Fano formulation.  Quasi-1D systems with different transverse confinement spectra, exhibit qualitative similarity in their resonance profiles, as is shown in Table. \ref{comparison}. This similarity comes from the mathematical structure of the transmission probability. All of them have the general form of the zeta function, namely
\begin{equation}
\xi_{H}(q,s)=\sum_{n=0}^{\infty}\dfrac{1}{(n+q)^{s}}
\end{equation}
is the Hurwitz zeta function that describes confinement-induced resonances in an isotropic transverse harmonic trap.  The quantity
\begin{equation}
\xi_{E}(q,s)=\sum_{n_x, n_y}^{\infty}\dfrac{1}{((n_x-q_x)^2+(n_y-q_y)^2+\epsilon)^{s}},
\end{equation}
is the Epstein zeta function, describing both the square well waveguide \cite{PhysRevA.88.012715} and the infinite 2D square lattice, in which $q=(q_x, q_y, \epsilon)$ denotes the ``shift\rq{}\rq{} coming from the ground state mode of system, the power $s$ is determined by the partial wave expansion of the interaction potential that is dominant in the relevant energy and symmetry of the system. For an $s$ wave, the zeta function parameter is $s=\frac{1}{2}$ \cite{Olshanii1998} and for a $p$ wave \cite{Blume2004}, $s=-\frac{1}{2}$. 
\par
The regularization method developed here works for both short range $s$ and $p$ wave scattering, and in principle for higher partial waves as well. The $d$-wave, \textit{e.g.}, is essential in describing the dipole-dipole interaction in a quasi-1D confinement geometry\cite{Giannakeas2012, arxiv1307.0899}. Mathematically, the Epstein zeta function and the Hurwitz zeta function have very different origins in analytic number theory. However, in describing the class of confinement-induced resonance phenomena, their emergence is quite intuitive and naturally generalized from one to another.
\par
%\begin{widetext}
%\onecolumngrid
\begin{table}[ht]
\caption{Comparison of confinement-induced resonances in different systems}
\centering
\begin{tabular}{cccl}
\includegraphics[scale=0.10]{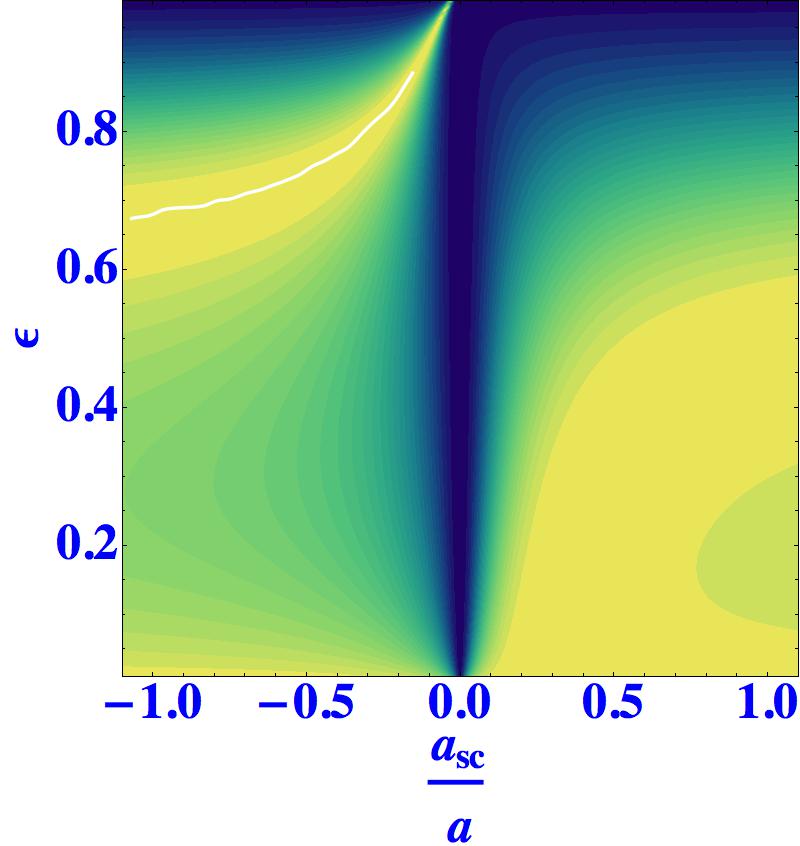} &
\includegraphics[scale=0.10]{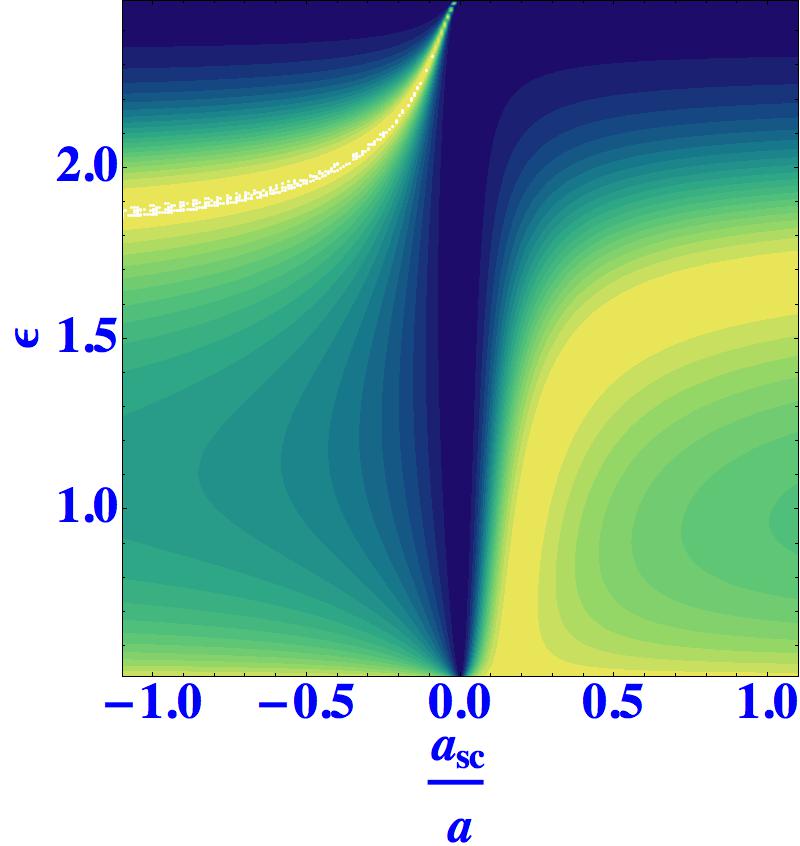} &
\includegraphics[scale=0.10]{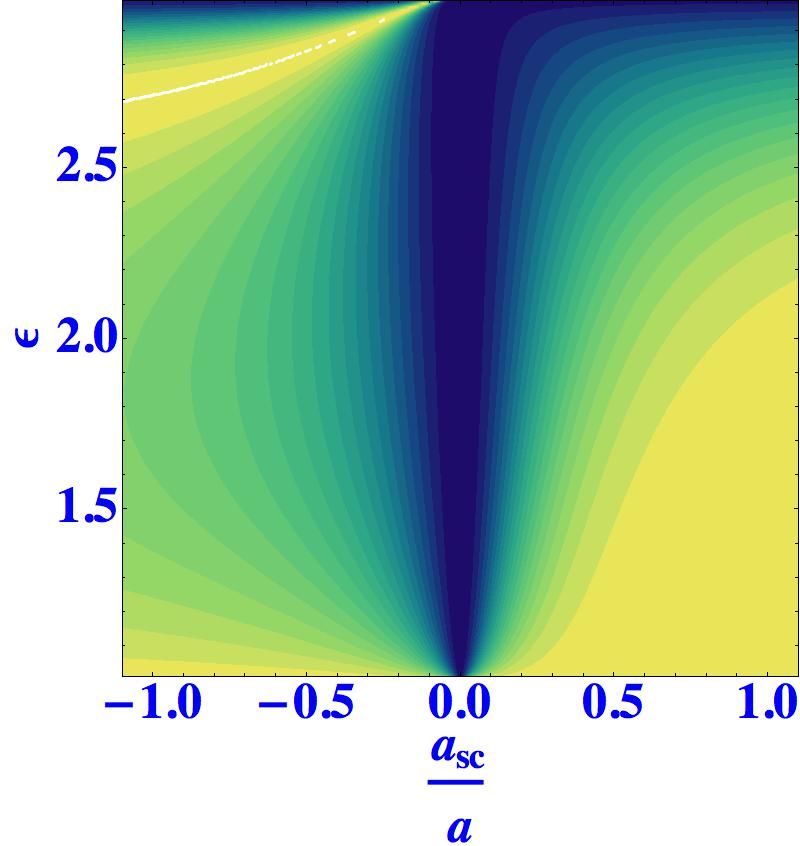} &
\includegraphics[scale=0.25]{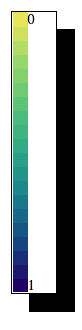}
\end{tabular}
(color online) The transmission probability for one open channel, in (from left to right) (i) two dimensional lattice (ii) square well waveguide (iii) two dimensional isotropic harmonic trap. The transmission probabilities for the case of one open channel are plotted. The units in each plot are the relevant system-characteristic energy scale, namely $\frac{\hbar^2}{2\mu}(\frac{2\pi}{a})^2$, $\frac{\hbar^2}{2\mu}(\frac{2\pi}{L})^2$, $\hbar\omega_{\perp}$. $a$ is the lattice constant, $L$ is the size of the square waveguide, and $\omega_\perp$ is the transverse trap frequency. Lighter colors correspond to less transmission, and the thin white line locates the positions of confinement-induced opacities.
\label{comparison}
\end{table}
%\end{widetext}
\par
In this Letter, the $s$-wave Feshbach resonances of a normally-incident matter wave scattered by atoms trapped in a 2D optical lattice are considered and shown to have an intimate connection with the 2D square confinement discussed previously \cite{PhysRevA.88.012715}. In future work, higher partial-wave Feshbach resonances between atoms in the matter wave and atoms trapped by the 2D optical lattice should be investigated. Although previous studies have considered a higher partial wave in a circular transverse trap configuration, the symmetry group of the 2D confinement is a subgroup of the symmetry group of the short range potential (mostly $O(3)$). This is not generally true for a higher partial-wave Feshbach resonance in an arbitrary 2D lattice configuration. Exploration of the quasi-1D scattering processes of $p$-waves, for instance, will give rise to a rich parameter space for realizing geometric phases.     
\par
In summary, a scheme for a matter wave transistor based on the resonant scattering process of an ultracold matter wave from a two dimensional optical lattice is proposed. Estimated temperatures have been given based on present experimental capabilities. Moreover, by utilizing the vanishing transmission coefficient of the 2D lattice plane for an incoming matter wave, it is possible to implement a matter wave cavity formed by two parallel 2D optical lattices. Finally, connections have been drawn between confinement-induced resonances in several quasi-1D systems. 
\par
We thank Doerte Blume, Yue Cao, Yong Chen, Abriham Olsen, J. P\'{e}rez-R\'{i}os, Yuan Sun, Desmond Yin, Hui Zhai for insightful discussions. This work is supported by NSF through grant Nos. PHY-1306905 and PHY-1125844. 
\appendix
%\section{Epstein zeta function}
%to be added


\begin{thebibliography}{34}
\expandafter\ifx\csname natexlab\endcsname\relax\def\natexlab#1{#1}\fi
\expandafter\ifx\csname bibnamefont\endcsname\relax
  \def\bibnamefont#1{#1}\fi
\expandafter\ifx\csname bibfnamefont\endcsname\relax
  \def\bibfnamefont#1{#1}\fi
\expandafter\ifx\csname citenamefont\endcsname\relax
  \def\citenamefont#1{#1}\fi
\expandafter\ifx\csname url\endcsname\relax
  \def\url#1{\texttt{#1}}\fi
\expandafter\ifx\csname urlprefix\endcsname\relax\def\urlprefix{URL }\fi
\providecommand{\bibinfo}[2]{#2}
\providecommand{\eprint}[2][]{\url{#2}}

\bibitem[{\citenamefont{Pepino et~al.}(2009)\citenamefont{Pepino, Cooper,
  Anderson, and Holland}}]{PhysRevLett.103.140405}
\bibinfo{author}{\bibfnamefont{R.~A.} \bibnamefont{Pepino}},
  \bibinfo{author}{\bibfnamefont{J.}~\bibnamefont{Cooper}},
  \bibinfo{author}{\bibfnamefont{D.~Z.} \bibnamefont{Anderson}},
  \bibnamefont{and} \bibinfo{author}{\bibfnamefont{M.~J.}
  \bibnamefont{Holland}}, \bibinfo{journal}{Phys. Rev. Lett.}
  \textbf{\bibinfo{volume}{103}}, \bibinfo{pages}{140405}
  (\bibinfo{year}{2009}).

\bibitem[{\citenamefont{Vaishnav et~al.}(2008)\citenamefont{Vaishnav, Ruseckas,
  Clark, and Juzeli\ifmmode~\bar{u}\else
  \={u}\fi{}nas}}]{PhysRevLett.101.265302}
\bibinfo{author}{\bibfnamefont{J.~Y.} \bibnamefont{Vaishnav}},
  \bibinfo{author}{\bibfnamefont{J.}~\bibnamefont{Ruseckas}},
  \bibinfo{author}{\bibfnamefont{C.~W.} \bibnamefont{Clark}}, \bibnamefont{and}
  \bibinfo{author}{\bibfnamefont{G.}~\bibnamefont{Juzeli\ifmmode~\bar{u}\else
  \={u}\fi{}nas}}, \bibinfo{journal}{Phys. Rev. Lett.}
  \textbf{\bibinfo{volume}{101}}, \bibinfo{pages}{265302}
  (\bibinfo{year}{2008}).

\bibitem[{\citenamefont{Micheli et~al.}(2004)\citenamefont{Micheli, Daley,
  Jaksch, and Zoller}}]{PhysRevLett.93.140408}
\bibinfo{author}{\bibfnamefont{A.}~\bibnamefont{Micheli}},
  \bibinfo{author}{\bibfnamefont{A.~J.} \bibnamefont{Daley}},
  \bibinfo{author}{\bibfnamefont{D.}~\bibnamefont{Jaksch}}, \bibnamefont{and}
  \bibinfo{author}{\bibfnamefont{P.}~\bibnamefont{Zoller}},
  \bibinfo{journal}{Phys. Rev. Lett.} \textbf{\bibinfo{volume}{93}},
  \bibinfo{pages}{140408} (\bibinfo{year}{2004}).

\bibitem[{\citenamefont{Qian et~al.}(2011)\citenamefont{Qian, Gong, and
  Zhang}}]{PhysRevA.84.013608}
\bibinfo{author}{\bibfnamefont{Y.}~\bibnamefont{Qian}},
  \bibinfo{author}{\bibfnamefont{M.}~\bibnamefont{Gong}}, \bibnamefont{and}
  \bibinfo{author}{\bibfnamefont{C.}~\bibnamefont{Zhang}},
  \bibinfo{journal}{Phys. Rev. A} \textbf{\bibinfo{volume}{84}},
  \bibinfo{pages}{013608} (\bibinfo{year}{2011}).

\bibitem[{\citenamefont{Riis et~al.}(1990)\citenamefont{Riis, Weiss, Moler, and
  Chu}}]{PhysRevLett.64.1658}
\bibinfo{author}{\bibfnamefont{E.}~\bibnamefont{Riis}},
  \bibinfo{author}{\bibfnamefont{D.~S.} \bibnamefont{Weiss}},
  \bibinfo{author}{\bibfnamefont{K.~A.} \bibnamefont{Moler}}, \bibnamefont{and}
  \bibinfo{author}{\bibfnamefont{S.}~\bibnamefont{Chu}},
  \bibinfo{journal}{Phys. Rev. Lett.} \textbf{\bibinfo{volume}{64}},
  \bibinfo{pages}{1658} (\bibinfo{year}{1990}).

\bibitem[{\citenamefont{Bloch et~al.}(1999)\citenamefont{Bloch, H\"ansch, and
  Esslinger}}]{PhysRevLett.82.3008}
\bibinfo{author}{\bibfnamefont{I.}~\bibnamefont{Bloch}},
  \bibinfo{author}{\bibfnamefont{T.~W.} \bibnamefont{H\"ansch}},
  \bibnamefont{and}
  \bibinfo{author}{\bibfnamefont{T.}~\bibnamefont{Esslinger}},
  \bibinfo{journal}{Phys. Rev. Lett.} \textbf{\bibinfo{volume}{82}},
  \bibinfo{pages}{3008} (\bibinfo{year}{1999}).

\bibitem[{\citenamefont{Mewes et~al.}(1997)\citenamefont{Mewes, Andrews, Kurn,
  Durfee, Townsend, and Ketterle}}]{PhysRevLett.78.582}
\bibinfo{author}{\bibfnamefont{M.-O.} \bibnamefont{Mewes}},
  \bibinfo{author}{\bibfnamefont{M.~R.} \bibnamefont{Andrews}},
  \bibinfo{author}{\bibfnamefont{D.~M.} \bibnamefont{Kurn}},
  \bibinfo{author}{\bibfnamefont{D.~S.} \bibnamefont{Durfee}},
  \bibinfo{author}{\bibfnamefont{C.~G.} \bibnamefont{Townsend}},
  \bibnamefont{and} \bibinfo{author}{\bibfnamefont{W.}~\bibnamefont{Ketterle}},
  \bibinfo{journal}{Phys. Rev. Lett.} \textbf{\bibinfo{volume}{78}},
  \bibinfo{pages}{582} (\bibinfo{year}{1997}).

\bibitem[{\citenamefont{Tomkovic et~al.}(2011)\citenamefont{Tomkovic,
  Schreiber, Welte, Kiffner, Schmiedmayer, and Oberthaler}}]{nature379382}
\bibinfo{author}{\bibfnamefont{J.}~\bibnamefont{Tomkovic}},
  \bibinfo{author}{\bibfnamefont{M.}~\bibnamefont{Schreiber}},
  \bibinfo{author}{\bibfnamefont{J.}~\bibnamefont{Welte}},
  \bibinfo{author}{\bibfnamefont{M.}~\bibnamefont{Kiffner}},
  \bibinfo{author}{\bibfnamefont{J.}~\bibnamefont{Schmiedmayer}},
  \bibnamefont{and} \bibinfo{author}{\bibfnamefont{M.~K.}
  \bibnamefont{Oberthaler}}, \bibinfo{journal}{Nature Phys.}
  \textbf{\bibinfo{volume}{7}}, \bibinfo{pages}{379} (\bibinfo{year}{2011}).

\bibitem[{\citenamefont{Caliga et~al.}(2012)\citenamefont{Caliga, Straatrma,
  Zozulya, and Anderson}}]{arxiv1208.3109}
\bibinfo{author}{\bibfnamefont{S.~C.} \bibnamefont{Caliga}},
  \bibinfo{author}{\bibfnamefont{C.~J.~E.} \bibnamefont{Straatrma}},
  \bibinfo{author}{\bibfnamefont{A.~A.} \bibnamefont{Zozulya}},
  \bibnamefont{and} \bibinfo{author}{\bibfnamefont{D.~Z.}
  \bibnamefont{Anderson}}, \bibinfo{journal}{arXiv}  (\bibinfo{year}{2012}),
  \eprint{1208.3109}.

\bibitem[{\citenamefont{Chen et~al.}(2013)\citenamefont{Chen, Beck, Bücker,
  Gullans, Lukin, Tanji-Suzuki, and Vuletic}}]{Chen16082013}
\bibinfo{author}{\bibfnamefont{W.}~\bibnamefont{Chen}},
  \bibinfo{author}{\bibfnamefont{K.~M.} \bibnamefont{Beck}},
  \bibinfo{author}{\bibfnamefont{R.}~\bibnamefont{Bücker}},
  \bibinfo{author}{\bibfnamefont{M.}~\bibnamefont{Gullans}},
  \bibinfo{author}{\bibfnamefont{M.~D.} \bibnamefont{Lukin}},
  \bibinfo{author}{\bibfnamefont{H.}~\bibnamefont{Tanji-Suzuki}},
  \bibnamefont{and} \bibinfo{author}{\bibfnamefont{V.}~\bibnamefont{Vuletic}},
  \bibinfo{journal}{Science} \textbf{\bibinfo{volume}{341}},
  \bibinfo{pages}{768} (\bibinfo{year}{2013}).

\bibitem[{\citenamefont{Chin et~al.}(2010)\citenamefont{Chin, Grimm, Julienne,
  and Tiesinga}}]{RevModPhys.82.1225}
\bibinfo{author}{\bibfnamefont{C.}~\bibnamefont{Chin}},
  \bibinfo{author}{\bibfnamefont{R.}~\bibnamefont{Grimm}},
  \bibinfo{author}{\bibfnamefont{P.}~\bibnamefont{Julienne}}, \bibnamefont{and}
  \bibinfo{author}{\bibfnamefont{E.}~\bibnamefont{Tiesinga}},
  \bibinfo{journal}{Rev. Mod. Phys.} \textbf{\bibinfo{volume}{82}},
  \bibinfo{pages}{1225} (\bibinfo{year}{2010}).

\bibitem[{\citenamefont{Kuklov and Svistunov}(1999)}]{PhysRevA.60.R769}
\bibinfo{author}{\bibfnamefont{A.~B.} \bibnamefont{Kuklov}} \bibnamefont{and}
  \bibinfo{author}{\bibfnamefont{B.~V.} \bibnamefont{Svistunov}},
  \bibinfo{journal}{Phys. Rev. A} \textbf{\bibinfo{volume}{60}},
  \bibinfo{pages}{R769} (\bibinfo{year}{1999}).

\bibitem[{\citenamefont{Sanders et~al.}(2010)\citenamefont{Sanders, Mintert,
  and Heller}}]{PhysRevLett.105.035301}
\bibinfo{author}{\bibfnamefont{S.~N.} \bibnamefont{Sanders}},
  \bibinfo{author}{\bibfnamefont{F.}~\bibnamefont{Mintert}}, \bibnamefont{and}
  \bibinfo{author}{\bibfnamefont{E.~J.} \bibnamefont{Heller}},
  \bibinfo{journal}{Phys. Rev. Lett.} \textbf{\bibinfo{volume}{105}},
  \bibinfo{pages}{035301} (\bibinfo{year}{2010}).

\bibitem[{\citenamefont{Cheiney et~al.}(2013)\citenamefont{Cheiney, Fabre,
  Vermersch, Gattobigio, Mathevet, Lahaye, and
  Gu\'ery-Odelin}}]{PhysRevA.87.013623}
\bibinfo{author}{\bibfnamefont{P.}~\bibnamefont{Cheiney}},
  \bibinfo{author}{\bibfnamefont{C.~M.} \bibnamefont{Fabre}},
  \bibinfo{author}{\bibfnamefont{F.}~\bibnamefont{Vermersch}},
  \bibinfo{author}{\bibfnamefont{G.~L.} \bibnamefont{Gattobigio}},
  \bibinfo{author}{\bibfnamefont{R.}~\bibnamefont{Mathevet}},
  \bibinfo{author}{\bibfnamefont{T.}~\bibnamefont{Lahaye}}, \bibnamefont{and}
  \bibinfo{author}{\bibfnamefont{D.}~\bibnamefont{Gu\'ery-Odelin}},
  \bibinfo{journal}{Phys. Rev. A} \textbf{\bibinfo{volume}{87}},
  \bibinfo{pages}{013623} (\bibinfo{year}{2013}).

\bibitem[{\citenamefont{Gadway et~al.}(2012)\citenamefont{Gadway, Pertot,
  Reeves, and Schneble}}]{Gadwaynp2012}
\bibinfo{author}{\bibfnamefont{B.}~\bibnamefont{Gadway}},
  \bibinfo{author}{\bibfnamefont{D.}~\bibnamefont{Pertot}},
  \bibinfo{author}{\bibfnamefont{J.}~\bibnamefont{Reeves}}, \bibnamefont{and}
  \bibinfo{author}{\bibfnamefont{D.}~\bibnamefont{Schneble}},
  \bibinfo{journal}{Nature Phys.} \textbf{\bibinfo{volume}{8}},
  \bibinfo{pages}{544} (\bibinfo{year}{2012}).

\bibitem[{\citenamefont{Stenger et~al.}(1999)\citenamefont{Stenger, Inouye,
  Chikkatur, Stamper-Kurn, Pritchard, and Ketterle}}]{PhysRevLett.82.4569}
\bibinfo{author}{\bibfnamefont{J.}~\bibnamefont{Stenger}},
  \bibinfo{author}{\bibfnamefont{S.}~\bibnamefont{Inouye}},
  \bibinfo{author}{\bibfnamefont{A.~P.} \bibnamefont{Chikkatur}},
  \bibinfo{author}{\bibfnamefont{D.~M.} \bibnamefont{Stamper-Kurn}},
  \bibinfo{author}{\bibfnamefont{D.~E.} \bibnamefont{Pritchard}},
  \bibnamefont{and} \bibinfo{author}{\bibfnamefont{W.}~\bibnamefont{Ketterle}},
  \bibinfo{journal}{Phys. Rev. Lett.} \textbf{\bibinfo{volume}{82}},
  \bibinfo{pages}{4569} (\bibinfo{year}{1999}).

\bibitem[{\citenamefont{Javanainen and
  Ruostekoski}(2003)}]{PhysRevLett.91.150404}
\bibinfo{author}{\bibfnamefont{J.}~\bibnamefont{Javanainen}} \bibnamefont{and}
  \bibinfo{author}{\bibfnamefont{J.}~\bibnamefont{Ruostekoski}},
  \bibinfo{journal}{Phys. Rev. Lett.} \textbf{\bibinfo{volume}{91}},
  \bibinfo{pages}{150404} (\bibinfo{year}{2003}).

\bibitem[{\citenamefont{Weitenberg et~al.}(2011)\citenamefont{Weitenberg,
  Schau\ss{}, Fukuhara, Cheneau, Endres, Bloch, and
  Kuhr}}]{PhysRevLett.106.215301}
\bibinfo{author}{\bibfnamefont{C.}~\bibnamefont{Weitenberg}},
  \bibinfo{author}{\bibfnamefont{P.}~\bibnamefont{Schau\ss{}}},
  \bibinfo{author}{\bibfnamefont{T.}~\bibnamefont{Fukuhara}},
  \bibinfo{author}{\bibfnamefont{M.}~\bibnamefont{Cheneau}},
  \bibinfo{author}{\bibfnamefont{M.}~\bibnamefont{Endres}},
  \bibinfo{author}{\bibfnamefont{I.}~\bibnamefont{Bloch}}, \bibnamefont{and}
  \bibinfo{author}{\bibfnamefont{S.}~\bibnamefont{Kuhr}},
  \bibinfo{journal}{Phys. Rev. Lett.} \textbf{\bibinfo{volume}{106}},
  \bibinfo{pages}{215301} (\bibinfo{year}{2011}).

\bibitem[{\citenamefont{Makotyn et~al.}(2013)\citenamefont{Makotyn, Klauss,
  Goldberger, Cornell, and Jin}}]{arxiv1308.3696}
\bibinfo{author}{\bibfnamefont{P.}~\bibnamefont{Makotyn}},
  \bibinfo{author}{\bibfnamefont{C.~E.} \bibnamefont{Klauss}},
  \bibinfo{author}{\bibfnamefont{D.~L.} \bibnamefont{Goldberger}},
  \bibinfo{author}{\bibfnamefont{E.~A.} \bibnamefont{Cornell}},
  \bibnamefont{and} \bibinfo{author}{\bibfnamefont{D.~S.} \bibnamefont{Jin}},
  \bibinfo{journal}{arxiv}  (\bibinfo{year}{2013}), \eprint{1308.3696}.

\bibitem[{\citenamefont{Regal et~al.}(2003)\citenamefont{Regal, Ticknor, Bohn,
  and Jin}}]{Regal2003}
\bibinfo{author}{\bibfnamefont{C.~A.} \bibnamefont{Regal}},
  \bibinfo{author}{\bibfnamefont{C.}~\bibnamefont{Ticknor}},
  \bibinfo{author}{\bibfnamefont{J.~L.} \bibnamefont{Bohn}}, \bibnamefont{and}
  \bibinfo{author}{\bibfnamefont{D.~S.} \bibnamefont{Jin}},
  \bibinfo{journal}{Nature} \textbf{\bibinfo{volume}{424}}, \bibinfo{pages}{47}
  (\bibinfo{year}{2003}).

\bibitem[{\citenamefont{Olshanii}(1998)}]{Olshanii1998}
\bibinfo{author}{\bibfnamefont{M.}~\bibnamefont{Olshanii}},
  \bibinfo{journal}{Phys. Rev. Lett.} \textbf{\bibinfo{volume}{81}},
  \bibinfo{pages}{938} (\bibinfo{year}{1998}).

\bibitem[{\citenamefont{Granger and Blume}(2004)}]{Blume2004}
\bibinfo{author}{\bibfnamefont{B.~E.} \bibnamefont{Granger}} \bibnamefont{and}
  \bibinfo{author}{\bibfnamefont{D.}~\bibnamefont{Blume}},
  \bibinfo{journal}{Phys. Rev. Lett.} \textbf{\bibinfo{volume}{92}},
  \bibinfo{pages}{133202} (\bibinfo{year}{2004}).

\bibitem[{\citenamefont{Kim et~al.}(2005)\citenamefont{Kim, Schmiedmayer, and
  Schmelcher}}]{PhysRevA.72.042711}
\bibinfo{author}{\bibfnamefont{J.~I.} \bibnamefont{Kim}},
  \bibinfo{author}{\bibfnamefont{J.}~\bibnamefont{Schmiedmayer}},
  \bibnamefont{and}
  \bibinfo{author}{\bibfnamefont{P.}~\bibnamefont{Schmelcher}},
  \bibinfo{journal}{Phys. Rev. A} \textbf{\bibinfo{volume}{72}},
  \bibinfo{pages}{042711} (\bibinfo{year}{2005}).

\bibitem[{\citenamefont{Zhang and Greene}(2013)}]{PhysRevA.88.012715}
\bibinfo{author}{\bibfnamefont{C.}~\bibnamefont{Zhang}} \bibnamefont{and}
  \bibinfo{author}{\bibfnamefont{C.~H.} \bibnamefont{Greene}},
  \bibinfo{journal}{Phys. Rev. A} \textbf{\bibinfo{volume}{88}},
  \bibinfo{pages}{012715} (\bibinfo{year}{2013}).

\bibitem[{\citenamefont{Fedichev et~al.}(2004)\citenamefont{Fedichev, Bijlsma,
  and Zoller}}]{PhysRevLett.92.080401}
\bibinfo{author}{\bibfnamefont{P.~O.} \bibnamefont{Fedichev}},
  \bibinfo{author}{\bibfnamefont{M.~J.} \bibnamefont{Bijlsma}},
  \bibnamefont{and} \bibinfo{author}{\bibfnamefont{P.}~\bibnamefont{Zoller}},
  \bibinfo{journal}{Phys. Rev. Lett.} \textbf{\bibinfo{volume}{92}},
  \bibinfo{pages}{080401} (\bibinfo{year}{2004}).

\bibitem[{\citenamefont{Elizalde}(1994)}]{e1994zeta}
\bibinfo{author}{\bibfnamefont{E.}~\bibnamefont{Elizalde}},
  \emph{\bibinfo{title}{Zeta Regularization Techniques With Applications}}
  (\bibinfo{publisher}{World Scientific Publishing Company, Incorporated},
  \bibinfo{year}{1994}), ISBN \bibinfo{isbn}{9789810214418}.

\bibitem[{\citenamefont{Botter et~al.}(2013)\citenamefont{Botter, Brooks,
  Schreppler, Brahms, and Stamper-Kurn}}]{PhysRevLett.110.153001}
\bibinfo{author}{\bibfnamefont{T.}~\bibnamefont{Botter}},
  \bibinfo{author}{\bibfnamefont{D.~W.~C.} \bibnamefont{Brooks}},
  \bibinfo{author}{\bibfnamefont{S.}~\bibnamefont{Schreppler}},
  \bibinfo{author}{\bibfnamefont{N.}~\bibnamefont{Brahms}}, \bibnamefont{and}
  \bibinfo{author}{\bibfnamefont{D.~M.} \bibnamefont{Stamper-Kurn}},
  \bibinfo{journal}{Phys. Rev. Lett.} \textbf{\bibinfo{volume}{110}},
  \bibinfo{pages}{153001} (\bibinfo{year}{2013}).

\bibitem[{\citenamefont{Romero-Isart et~al.}(2013)\citenamefont{Romero-Isart,
  Navau, Sanchez, Zoller, and Cirac}}]{PhysRevLett.111.145304}
\bibinfo{author}{\bibfnamefont{O.}~\bibnamefont{Romero-Isart}},
  \bibinfo{author}{\bibfnamefont{C.}~\bibnamefont{Navau}},
  \bibinfo{author}{\bibfnamefont{A.}~\bibnamefont{Sanchez}},
  \bibinfo{author}{\bibfnamefont{P.}~\bibnamefont{Zoller}}, \bibnamefont{and}
  \bibinfo{author}{\bibfnamefont{J.~I.} \bibnamefont{Cirac}},
  \bibinfo{journal}{Phys. Rev. Lett.} \textbf{\bibinfo{volume}{111}},
  \bibinfo{pages}{145304} (\bibinfo{year}{2013}).

\bibitem[{\citenamefont{Fano}(1961)}]{PhysRev.124.1866}
\bibinfo{author}{\bibfnamefont{U.}~\bibnamefont{Fano}}, \bibinfo{journal}{Phys.
  Rev.} \textbf{\bibinfo{volume}{124}}, \bibinfo{pages}{1866}
  (\bibinfo{year}{1961}).

\bibitem[{\citenamefont{Gupta et~al.}(2001)\citenamefont{Gupta, Leanhardt, D.,
  and E.}}]{Guptapaper}
\bibinfo{author}{\bibfnamefont{S.}~\bibnamefont{Gupta}},
  \bibinfo{author}{\bibfnamefont{A.~E.} \bibnamefont{Leanhardt}},
  \bibinfo{author}{\bibfnamefont{C.~A.} \bibnamefont{D.}}, \bibnamefont{and}
  \bibinfo{author}{\bibfnamefont{P.~D.} \bibnamefont{E.}}, \bibinfo{journal}{C.
  R. Acac. Sci. Ser IV: Phys., Astrophys.} \textbf{\bibinfo{volume}{2}},
  \bibinfo{pages}{479} (\bibinfo{year}{2001}).

\bibitem[{\citenamefont{Poli et~al.}(2011)\citenamefont{Poli, Wang, Tarallo,
  Alberti, Prevedelli, and Tino}}]{PhysRevLett.106.038501}
\bibinfo{author}{\bibfnamefont{N.}~\bibnamefont{Poli}},
  \bibinfo{author}{\bibfnamefont{F.-Y.} \bibnamefont{Wang}},
  \bibinfo{author}{\bibfnamefont{M.~G.} \bibnamefont{Tarallo}},
  \bibinfo{author}{\bibfnamefont{A.}~\bibnamefont{Alberti}},
  \bibinfo{author}{\bibfnamefont{M.}~\bibnamefont{Prevedelli}},
  \bibnamefont{and} \bibinfo{author}{\bibfnamefont{G.~M.} \bibnamefont{Tino}},
  \bibinfo{journal}{Phys. Rev. Lett.} \textbf{\bibinfo{volume}{106}},
  \bibinfo{pages}{038501} (\bibinfo{year}{2011}).

\bibitem[{\citenamefont{Charri\`ere et~al.}(2012)\citenamefont{Charri\`ere,
  Cadoret, Zahzam, Bidel, and Bresson}}]{PhysRevA.85.013639}
\bibinfo{author}{\bibfnamefont{R.}~\bibnamefont{Charri\`ere}},
  \bibinfo{author}{\bibfnamefont{M.}~\bibnamefont{Cadoret}},
  \bibinfo{author}{\bibfnamefont{N.}~\bibnamefont{Zahzam}},
  \bibinfo{author}{\bibfnamefont{Y.}~\bibnamefont{Bidel}}, \bibnamefont{and}
  \bibinfo{author}{\bibfnamefont{A.}~\bibnamefont{Bresson}},
  \bibinfo{journal}{Phys. Rev. A} \textbf{\bibinfo{volume}{85}},
  \bibinfo{pages}{013639} (\bibinfo{year}{2012}).

\bibitem[{\citenamefont{Giannakeas et~al.}(2012)\citenamefont{Giannakeas,
  Diakonos, and Schmelcher}}]{Giannakeas2012}
\bibinfo{author}{\bibfnamefont{P.}~\bibnamefont{Giannakeas}},
  \bibinfo{author}{\bibfnamefont{F.~K.} \bibnamefont{Diakonos}},
  \bibnamefont{and}
  \bibinfo{author}{\bibfnamefont{P.}~\bibnamefont{Schmelcher}},
  \bibinfo{journal}{Phys. Rev. A} \textbf{\bibinfo{volume}{86}},
  \bibinfo{pages}{042703} (\bibinfo{year}{2012}).

\bibitem[{\citenamefont{Guan et~al.}(2013)\citenamefont{Guan, Cui, Qi, and
  Zhai}}]{arxiv1307.0899}
\bibinfo{author}{\bibfnamefont{L.}~\bibnamefont{Guan}},
  \bibinfo{author}{\bibfnamefont{X.}~\bibnamefont{Cui}},
  \bibinfo{author}{\bibfnamefont{R.}~\bibnamefont{Qi}}, \bibnamefont{and}
  \bibinfo{author}{\bibfnamefont{H.}~\bibnamefont{Zhai}},
  \bibinfo{journal}{arxiv}  (\bibinfo{year}{2013}), \eprint{1307.0899}.

\end{thebibliography}
\end{document}